\documentclass[english,english,aps,prd,amsmath,amssymb,superscriptaddress]{revtex4}
\usepackage[T1]{fontenc}
\usepackage[latin9]{inputenc}
\usepackage{amsmath}
\usepackage{amssymb}
\usepackage{esint}

\makeatletter
\@ifundefined{textcolor}{}
{%
 \definecolor{BLACK}{gray}{0}
 \definecolor{WHITE}{gray}{1}
 \definecolor{RED}{rgb}{1,0,0}
 \definecolor{GREEN}{rgb}{0,1,0}
 \definecolor{BLUE}{rgb}{0,0,1}
 \definecolor{CYAN}{cmyk}{1,0,0,0}
 \definecolor{MAGENTA}{cmyk}{0,1,0,0}
 \definecolor{YELLOW}{cmyk}{0,0,1,0}
 }

\usepackage{slashed}
\usepackage{babel}
\usepackage{bbm}

\makeatother

\usepackage{babel}

\begin{document}

\title{The Cosmological Constant Problem and Re-interpretation of Time}

\author{M.J.Luo}

\affiliation{Department of Physics, Tsinghua University, Beijing 100084, People's
Republic of China}
\begin{abstract}
We abandon the interpretation that time is a global parameter in quantum
mechanics, replace it by a quantum dynamical variable playing the
role of time. This operational re-interpretation of time provides
a solution to the cosmological constant problem. The expectation value
of the zero-point energy under the new time variable vanishes. The
fluctuation of the vacuum energy as the leading contribution to the
gravitational effect gives a correct order to the observed {}``dark
energy''. The {}``dark energy'' as a mirage is always seen comparable
with the matter energy density by an observer using the internal clock
time. Conceptual consequences of the re-interpretation of time are
also discussed.
\end{abstract}
\maketitle

\section{Introduction}

The cosmological constant problem is a crisis of physics \cite{RevModPhys.61.1,RevModPhys.75.559}.
It arises as a severe problem because anything that contributes to
the energy density of the vacuum behaves like a cosmological constant.
For example, contributions come from the potential of scalar Higgs
boson, which is about $(200\mathrm{GeV})^{4}$; the chiral symmetry
breaking of QCD is about $(300\mathrm{MeV})^{4}$; the behavior of
electrons is well understood up to energies of order $100\mathrm{GeV}$,
so the contribution of electron loops up to this scale contribute
of order $(100\mathrm{GeV})^{4}$ to the vacuum energy; and other
conjectured scale, like the supersymmetry breaking scale gives at
least $(1\mathrm{TeV})^{4}$ to the vacuum energy. Among various known
contributions, the most severe trouble comes from the so-called zero-point
vacuum energy predicted from our well-tested quantum field theory.
We know from the quantum field theory that our vacuum is rather non-trivial,
the sum of the zero-point energies of all normal modes of quantum
fields gives $\sum_{k}^{k_{max}}\frac{1}{2}\hbar\omega_{k}\approx\frac{k_{max}^{4}}{16\pi^{2}}$.
If we believe the general relativity up to the Planck scale $k_{max}\sim10^{19}\mathrm{GeV}$,
it would give about $(10^{19}\mathrm{GeV})^{4}$. However, the present
observations measure the effective vacuum energy and give a very small
value, about $(10^{-12}\mathrm{GeV})^{4}$ \cite{Riess:1998cb,2013arXiv1303.5062P}.

It is very disappointing for this large difference (about $10^{120}$)
from the prediction in quantum field theory. This would need to be
canceled almost, but no exactly, by an equally large counter term
with opposite sign, using the standard renormalization procedure of
the quantum field theory. The cancellation of the quartic divergence
and large magnitude of quantum correction compared with its small
bare value leads to a severe fine-tuning. We have no reason why these
large amount of quantum corrections of vacuum do not gravitate. If
we trust the well-tested equivalence principle of the general relativity,
all kinds of energies gravitate. Indeed, we know that the electron
vacuum energy coming from the vacuum polarization (measured by the
famous Lamb shift experiment) does gravitate \cite{2006hep.th....3249P}.
Explaining why these quantum corrections do not gravitate is only
one side, why they leave a small remnant gravitational effect is another
side, since some supersymmetric theories really require an exact zero
vacuum energy (although it does not help because the supersymmetry
must be broken). Current cosmological observation raises the third
part of the cosmological constant problem called the cosmic coincidence
problem or {}``why now'' problem \cite{2008GReGr..40..607B}, i.e.
why it is comparable to the matter energy density now. Because the
vacuum energy or anything behaving like it does not redshift like
matter. In the past, the matter density is large, and the vacuum energy
can be ignored, but in the latter time, matter will gradually be diluted
by the expansion of the universe, but the vacuum energy density remains,
so the percentage of the vacuum energy become large. These two can
be comparable only in a particular epoch, but we have no idea why
it is now?

What the cosmological constant problem actually implies for our most
fundamental concepts and understanding of the world is not yet clear.
But one thing is clear, if one wish to talk about the notion of energy,
one must bear in mind very carefully that a mathematical description
of it has no physical meaning unless one really clears what the {}``time''
means \cite{PhysRevD.42.2638,rovelli1991time,rovelli2004quantum,Rovelli:2009ee}.
In its usual sense, if the notion of time is defined as {}``the position
of the pointer of the clock in my hand'', we place this clock as
close as the system, and the energy exists as an abstract and mysterious
mathematical quantity that you find it always the same under the position
change of the clock pointer. This is true, but not true enough. In
certain situations, the position of the clock pointer becomes fuzzy.
Beside that, if the system is very far from us, and the clock by definition
is placed far apart as well, certain technique is needed to compare
the readings of it and that of the clock in my hand, since there is
no further assumption to compare how fast these two clocks are, even
if they have already been locally synchronized. To measure the energy
at a distance, we have to take into account that all our judgments
in which energy or time concerns are always the judgments of the synchronization
of clocks in a distance. However, can the synchronization between
spatially separated clocks be precisely realized? The answer from
our current understanding of the spacetime based on the classical
relativity is {}``yes''. But what is the case when the spacetime
is quantum mechanical? In principle, this kind of question cannot
be fully answered unless a consistent quantum theory of spacetime
is discovered. But it is because the lacking of the complete theory
of quantum spacetime, the cosmological constant problem may be an
important clue to find the theory, the notion of time may be a key
to the problem.

Before we discuss the problem, let's first re-examine what the concept
of time actually is in section II, in which we re-interpret the notion
of time and present our framework. Then we discuss the consequences
of the re-interpretation of time and its implication to the cosmological
constant problem in section III. In section IV, we generalize the
idea to a more general version of quantum reference frame. Finally,
we draw the conclusions of the paper in section V.

\section{Operational Definition of Time at Quantum Level}

The modern physics from his father Galileo and Newton was built beginning
with the physical realization of time. Galileo gave a decisive contribution
to the discovery of the modern clock. The small oscillations of a
pendulum can be used as a standard clock that {}``takes equal time'',
although he had no standard {}``time'' to tell him whether its oscillation
periods take equal time, he just checked the pendulum against his
pulse (which was thought as the only {}``standard'' he could use).
It was Newton who made the conceptual idea clear up. He assumed the
existence of a global parameter measured by the device invented by
Galileo, which flows to infinity absolutely as the pendulum oscillates
forever. By using this global parameter ({}``time''), the law of
motion could be simplified. The observable quantities can be parametrized
as functions of Newton's time $X(t),Y(t),Z(t)$..., called evolution.
The complex motions we observed in nature are strongly simplified
by a few fundamental laws that governs the form of these functions.
The physics was then built up to predict the behavior of $X(t),Y(t),Z(t)$
at temporal distance, in analog with the Euclidean geometry (before
Newton's mechanics thousand of years) which predicts the behavior
of points, lines and angles at spatial distance. This framework of
the universe was challenged when the law of electrodynamics discovered
experimentally was found conflict with the relativity of Galileo.
It was Einstein who cleared up the issue conceptually that we must
re-examine the simultaneity in different reference frames. He found
that the clock readings (by his invention of {}``light clock'')
in different reference frames are not Newton's global parameter $t$
but each reference frame has each parameter time, as a consequence
of the constant speed of light. The physical quantities can be rewritten
as $X(\tau),Y(\tau),Z(\tau)$... together with his light clock $T(\tau)$,
if a global parameter $\tau$ is also assumed to be exist. He abandoned
the unobserved global time, and replaced it by physical clocks readings
from his light clock in each reference frame. Then the evolution in
each reference frame is seen as functionals $X[T(\tau)],Y[T(\tau)],Z[T(\tau)]$...
instead of the functions $X(t),Y(t),Z(t)$.... 

Newton's global parameter in certain sense is still alive in the quantum
mechanics even the quantum fields theories after the discovery of
Einstein's re-interpretation of time. Heisenberg abandoned the unobserved
spatial trajectories of electron in atom, and only used the observable
such as light spectrum that induced from the transition between two
states. As a consequence the spatial coordinate of electron as a number
was replaced by a square matrix relating two states. Although the
spatial coordinates was re-interpreted, the time in quantum mechanics
is still the classical Newton's parameter. The quantum fields theories
replaced the only one global parameter in quantum mechanics by four
interpreted as the spacetime coordinates in order to keep the Lorentz
invariance. The quantum mechanics presupposes an external classical
observer measuring the global parameter time, which makes several
intrinsic difficulties, e.g. the quantum mechanics cannot be applied
to study the whole universe, since the universe has no outside by
its definition. And such division of the universe into a quantum world
that to-be-measured and a classical world describing the measuring
instruments, makes the quantum mechanics needs extra assumptions or
axioms to justify the process of measurement, such as the argument
of the collapse of wavefunction in the Copenhagen interpretation of
quantum mechanics. 

As a general believe, the difficulty of quantizing general relativity
is deeply rooted in the very different treatment of the concept of
time in general relativity and in quantum mechanics. The lessons we
have learned about our world from these two theories is that we need
to carefully reexamine our fundamental notion. The relativity teaches
us that the time is nothing but an artificial notion invented to simplify
our thinking about motion, in classical physics the clock is always
imagined as an idea or perfect motion that as a reference to other
more complex motions. While the quantum mechanics teaches us that
there is no idea or perfect motion in our world, the physical quantities
are always fluctuating quantum mechanically. However, the non-existence
of a perfect motion as a standard clock in quantum mechanics may not
be important, in practice, all clocks including the quantum or atom
clocks we have invented in laboratories are not perfect, they are
just used as reference like the relativity had taught us, the important
thing is the relation between different motions, they could be bridged
by an imagined perfect motion or not by it, whether the bridge exists
or not the relations are still there, just like that we still could
exchange our goods without money.

A physical theory works in such a way, a formal mathematical apparatus
joints with a physical interpretation. The great progresses in the
history of physics are made not so much through a deeper understanding
of the nature as a deeper understanding of the science itself. Combining
the spirits of the relativity and quantum mechanics, the physical
clocks time $T(\tau)$ used in the relativity must be treated quantum
mechanically, and then the global parameter $\tau$ in quantum mechanics
can no longer be interpreted as time. This re-interpreted time variable
is relativistic, since it is physical operational defined; it is quantum,
since the physical clock $T(\tau)$ is treated quantum mechanically
and has quantum fluctuation. As a key new ingredient, the classical
relativistic simultaneity cannot be realized precisely due to the
intrinsic quantum fluctuation, just like Newton's simultaneity cannot
be precisely realized when the speed of light is a constant.

What does it mean when we consider the physical clock $T(\tau)$ is
quantum mechanical, and what is the meaning when we talk about the
quantum version of the evolution $X[T(\tau)],Y[T(\tau)],Z[T(\tau)]$...
Let us consider a Hamiltonian $H_{X}$ governing the behavior of the
physical quantities $X(\tau)$, and a Hamiltonian $H_{T}$ governing
the physical clock $T(\tau)$. They share the global parameter $\tau$
in quantum mechanical treatment, whether or not $\tau$ has any physical
meanings is not important in our setting. There is no interaction
between the field $X(\tau)$ and $T(\tau)$, it is a separable system,
each field independently evolves with the parameter $\tau$, so the
Hilbert space of the system is a direct product of these two Hilbert
spaces $\mathcal{H}=\mathcal{H}_{X}\otimes\mathcal{H}_{T}$, the state
vector can be written as $\left|\Psi\right\rangle =\sum_{\tau}c_{\tau}\left|X(\tau)\right\rangle \otimes\left|T(\tau)\right\rangle $
which is the eigenstate of $H=H_{X}+H_{T}$. The statement that the
system is separable does not necessarily mean that they are always
independent, since when we initialize an experiment we need to adjust
the instruments, which makes an instant interacting between $X(\tau)$
and $T(\tau)$ at early stage, and hence the state $\left|\Psi\right\rangle $
is not simply a direct product state $\sum_{\tau}a_{\tau}\left|X(\tau)\right\rangle \otimes\sum_{\tau^{\prime}}b_{\tau^{\prime}}\left|T(\tau^{\prime})\right\rangle $,
in most cases, it is an entangled state. This argument suggest that
the evolution $X[T(\tau)]$, at quantum level, is replaced by the
entangled state $\sum_{\tau}c_{\tau}\left|X(\tau)\right\rangle \otimes\left|T(\tau)\right\rangle $.
The squared norm of the coefficient of the entangled state $\left|c_{\tau}\right|^{2}$
measures the joint probability when the clock is at state $\left|T(\tau)\right\rangle $
and the physical quantity is at $\left|X(\tau)\right\rangle $, which
is a quantum version of the process that one reads the clock and sees
the evolution of $X$. Only when the clock is classical and deterministic,
$\left|c_{\tau}\right|^{2}$ reduces to the textbook probability of
$\left|X(\tau)\right\rangle $. The relational probabilistic interpretation
of $\left|X(\tau)\right\rangle \otimes\left|T(\tau)\right\rangle $
replaces the deterministic interpretation of $X[T(\tau)]$, and the
Schrodinger equation governing the quantum evolution of $\left|X(\tau)\right\rangle $
and $\left|T(\tau)\right\rangle $ with $\tau$ is replaced by the
Wheeler-DeWitt equation \begin{equation}
\left(H_{X}+H_{T}\right)\left|\Psi\right\rangle =\left(H_{X}+H_{T}\right)\sum_{\tau}c_{\tau}\left|X(\tau)\right\rangle \otimes\left|T(\tau)\right\rangle =0.\label{eq:wheeler-dewitt}\end{equation}
The quantum physical clock time $T(\tau)$ reduces to the classical
relativistic time when we use the mean field approximation. The Schrodinger
equation emerges as an approximation from the above Wheeler-DeWitt
equation.

Consider the action of the separable system $S=S_{X}+S_{T}$, where
the physical clock by convention is chosen as the quantum pendulum,
i.e. the continuous free quantum fields or infinite many quantum harmonic
oscillators located on the parameter background $\tau$, $S_{T}=\frac{1}{2}\int d^{d}\tau\left(\partial_{\tau}T\right)^{2}$,
the action of $X$ can be in general written as conventional kinetic
part and potential energy part $S_{X}=\int d^{d}\tau\frac{1}{2}\left(\partial_{\tau}X\right)^{2}-V[X(\tau)]$.
The action can be written as\begin{align}
S & =\int d^{d}\tau\left[\frac{1}{2}\left(\partial_{\tau}X\right)^{2}-V[X]+\frac{1}{2}\left(\partial_{\tau}T\right)^{2}\right]\label{eq:action-matter+clock}\\
 & =\int dT\left\Vert \frac{\partial\tau}{\partial T}\right\Vert \left[\frac{1}{2}\left(\partial_{\tau}T\right)^{2}\left[1+\left(\frac{\delta X}{\delta T}\right)^{2}\right]-V[X]\right],\label{eq:action-matter+clock-variation}\end{align}
where $d$ is the dimension of the parameter space $\tau$, the $\left\Vert \frac{\partial\tau}{\partial T}\right\Vert $
is the Jacobian determinant. Then the partition function is\begin{equation}
Z=\int\mathcal{D}X\mathcal{D}T\exp\left(-S[X(\tau),T(\tau)]\right)\overset{MF}{\approx}\int\mathcal{D}X\exp\left(-S_{eff}[X[T]]\right),\end{equation}
where the effective action under the mean field approximation is \begin{equation}
S_{eff}\left[X,\frac{\delta X}{\delta T}\right]=\int dT\frac{1}{2}\mathcal{M}\left(\frac{\delta X}{\delta T}\right)^{2}-V[X]+\mathrm{const},\label{eq:effective-action-matter+clock}\end{equation}
where $\mathcal{M}=\left\langle \left\Vert \frac{\partial\tau}{\partial T}\right\Vert \left(\partial_{\tau}T\right)^{2}\right\rangle _{MF}$
is a constant depending on the integration constant of the mean field
value of $T(\tau)$. Up to a constant, the mean field effective action
reproduces the classical action of $X$, $S=\int dt\frac{1}{2}m\left(\frac{dx}{dt}\right)^{2}-V(x)$,
by using the $T$ as time of the system. An obvious observation from
this effective action is that, the functional derivative formally
replaces the conventional derivative, since the clock time $T(\tau)$
now is certain dynamical variable playing the role of time. The evolution
of $X$ is now with respect to the physical quantity $T$.

The notion of time and energy are closely correlated to each other,
this is true not only in quantum mechanics, but also in classical
physics. The textbook Schrodinger equation strongly relies on the
notion of energy, but in our setting, strictly speaking there is no
notion of time and energy at fundamental level. Only when the quantum
fluctuated clock time $T(\tau)$ is treated semi-classically as a
classical parameter time, the Schrodinger equation emerges as an approximation
from the Wheeler-DeWitt equation, and the conventional notion of {}``time''
in the induced Schrodinger equation emerges from the timeless Wheeler-DeWitt
equation. The effective action Eq.(\ref{eq:effective-action-matter+clock})
and the emergent Schrodinger equation are only approximations, as
a consequence, the notion of unitarity of the Schrodinger equation
is also an approximation. That is not to say that the probability
does not conserve any more, it suggests that certain relational interpretation
connected to the entangled state solution of the Wheeler-DeWitt equation
must be introduced, replacing the absolute probability in the textbook
quantum mechanics. The relational interpretation just cares about
the mutual relation between the to-be-measured system $\left|X(\tau)\right\rangle $
and measuring instrument (the clock) $\left|T(\tau)\right\rangle $,
the absolute individual state $\left|X(\tau)\right\rangle $ or $\left|T(\tau)\right\rangle $
defined as a function of $\tau$ has no individual physical meaning.
In the standard Schrodinger picture, a state is defined as a function
of the parameter time, the entangled state of the Wheeler-DeWitt equation
can also be defined as a function of its global parameter, but it
is not interpreted as time any more. The state of the to-be-measured
system is defined (to be) entangled with the state of the clock, it
is their relation that are observable. It is reasonable that only
the relation between the to-be-measured system and the measuring instruments
is important, but their individual absolute states, when the measuring
instrument such as the clock is inevitably be treated quantum mechanically. 

The action Eq.(\ref{eq:action-matter+clock}) and the related Wheeler-DeWitt
equation Eq.(\ref{eq:wheeler-dewitt}) are the precise theories we
need to study quantum mechanically. We hence face a boundary of the
textbook Schrodinger equation and beyond, in this sense, the functional
approach is more useful than the operator approach. Beyond the mean
field approximation, the quantum fluctuations of the clock time become
important, which will lead to a departure from the prediction of the
parameter time. A salient departure gives rise to the first order
prediction of the groundstate energy, which exhibits the importance
of the quantum fluctuation of the clock, the classical relativistic
simultaneity cannot by precisely realized due to the intrinsic quantum
fluctuation. These consequences will be discussed in detail in the
next section.

\section{Consequences-The Cosmological Constant Problem}

In this section, we will discuss the consequences of the operational
re-interpreted time, which leads to a solution of the cosmological
constant problem. True enough, as a fundamental variable of physics,
any modification of the notion of time will cause salient changes
and consequences in almost all aspects of physics. The most directly
related notion is no doubt the energy. 

Considering the whole universe is divided into two sub-regions, the
finite region-1 with a sphere of radius $R$, and the outside region,
denoted as region-2. And considering a dynamical system with a total
action be written as $S[X_{1},X_{2}]=S_{1}[X_{1}]+S_{2}[X_{2}]$,
where $S_{1}[X_{1}]$ is the action of the system obtained by integrating
the Lagrangian density within region-1 and $S_{2}$ is the action
of the outside region, region-2. The $X_{1}(x)$ and $X_{2}(x)$ are
quantum scalar field variables in region-1 and region-2 respectively,
$x$ are global parameters shared by these fields. These two fields
independently live in each region and do not couple with each other.
Since there is no external classical observer outside the whole region
(outside the region-1 and region-2), the parameter time $t=x_{0}$
cannot be interpreted any more as the notion of time with any physical
meaning, the total energy of the whole system cannot be observed. 

In such a closed quantum system without outside, measuring a subsystem
means that an observer stands outside the subsystem and uses measuring
instruments to {}``watch'' it from the external of the subsystem.
Now let us consider an observer in the region-1 performs measurements
to {}``watch'' the system of the region-2, by using a device described
by a field $X_{1}$ being a clock in his/her hands. The energy of
the region-2 can only be measured by the observer's clock readings
$X_{1}$, which is a quantity defined as his/her clock time shift
invariant. The energy density of region-2 is considered to be continued
to the point $x$ when the region-1 tends to shrink to the point $x$,\begin{equation}
\left\langle E_{2}(x)\right\rangle =\frac{\delta S_{eff}}{\delta X_{1}(x)},\end{equation}
where the effective action is $S_{eff}=-\ln Z=-\ln\int\mathcal{D}X_{1}\mathcal{D}X_{2}e^{-(S_{1}+S_{2})}$.
Here the functional derivative w.r.t. the clock time fields replaces
the conventional derivative w.r.t. the global parameter time in defining
the energy $E=\frac{\partial S}{\partial t}$. The latter global energy
can only be measured by an external observer outside the whole system,
so it is completely unobservable in our setting. 

In principle, a clock is just a reference, any dynamical variable
could be chosen and defined as a clock, but by convention, the simplest
and practical ones are the periodic systems. Now let us consider the
physical clock $X_{1}$ is the coordinate of a periodic quantum harmonic
oscillator or a continuous free quantum field (infinite many quantum
harmonic oscillators located at the continuous parameter $x$), the
action of the physical clock is written as\begin{equation}
S_{1}[X_{1}]=\frac{1}{2}\int_{\left|x\right|<R}d^{4}x\left(\partial_{x}X_{1}\right)^{2}.\label{eq:action-clock}\end{equation}

The first consequence of the framework is that the vacuum energy of
$E_{2}$ measured by the observer in region-1 is vanished. Note that
the subsystem $S_{1}$ and $S_{2}$ are independent, i.e. only $S_{1}$
contains explicitly the field variable $X_{1}$, so in fact the energy
$E_{2}$ density is just the conjugate momentum density $p_{1}$ of
the field $X_{1}$,\begin{equation}
\left\langle E_{2}\right\rangle =\frac{\delta S_{eff}}{\delta X_{1}}=\left\langle p_{1}\right\rangle .\end{equation}
Let $\left|0\right\rangle =\left|0\right\rangle _{1}\otimes\left|0\right\rangle _{2}$
be the groundstate which is the eigenstate of the Hamiltonian of the
whole system. It is known that the expectation value of momentum $p_{1}$
at groundstate is trivially vanished $\left\langle 0\right|p_{1}\left|0\right\rangle =0$,
so we find \begin{equation}
\left\langle 0\right|E_{2}\left|0\right\rangle =0.\end{equation}
This result exhibits that the observer does not feel any zero-point
energy density of the system of the region-2. The divergent zero-point
energy density predicted by conventional quantum field theories $\frac{1}{2}\hbar\sum_{\omega}\omega$
can only be seen by an external classical observer outside the universe
using the global parameter time. Only the energy of a subsystem can
be measured, one needs to stand outside the subsystem, and use the
physically fluctuating field variable outside the subsystem as the
physical clock time. When the clock $X_{1}$ in region-1 is treated
quantum mechanically, it also undergoes zero-point fluctuation. As
a consequence, we cannot feel the zero-point fluctuations by using
a zero-point fluctuating clock, or equivalently standing on a zero-point
fluctuating reference frame. In summary, we abandon the unobserved
parameter time and use the operational defined quantum clock variable
as time, the zero-point energy automatically vanishes, then there
are no such divergent contributions to the cosmological constant. 

The second consequence from the re-interpretation of time is that
when we consider the quantum fluctuations of the physical clock time,
the intrinsic quantum uncertainty in the notion of simultaneity between
two clocks will result in intrinsic quantum fluctuation of energy
density, leading to an observed order of the mysterious energy density
(so-called the {}``dark energy'') that drives the accelerating expansion
of the universe. In our setting, the effective energy density is completely
due to the quantum effects of the re-interpreted time variable, and
there is no need for the extra assumption of dark energy. To deduce
such consequence, note that although $\left\langle E_{2}\right\rangle =\frac{\delta S}{\delta X_{1}}=0$,
we have a non-vanished zero-point energy fluctuation $\left\langle \delta E_{2}^{2}\right\rangle =\left\langle E_{2}^{2}\right\rangle -\left\langle E_{2}\right\rangle ^{2}=\left\langle E_{2}^{2}\right\rangle =\frac{\delta^{2}S}{\delta X_{1}^{2}}\neq0$.

The zero-point energy fluctuation can be understood as follows. Considering
the physical clocks $X_{1}$ at spatially separated points $x$ and
$y$, with clocks readings $X_{1}(x)$ and $X_{1}(y)$. To compare
these two clocks quantum mechanically, a quantitative description
is by a probabilistic correlation function, from Eq.(\ref{eq:action-clock})
we have\begin{equation}
\left\langle X_{1}(x)X_{1}(y)\right\rangle =\int\frac{d^{4}k}{(2\pi)^{4}}\frac{1}{k^{2}}e^{ik\cdot(x-y)}\sim\frac{1}{4\pi^{2}\left|x-y\right|^{2}},\label{eq:clock-correlation}\end{equation}
which measures the correlation between the clock at $x$ reads $X_{1}(x)$
and at $y$ reads $X_{1}(y)$. Note that the correlation between the
two clocks decays with their spatial distance. The decorrelation between
the clocks indicates the fact that the rates of these two clocks are
unable to be synchronized precisely at quantum level. There is an
intrinsic uncertainty in synchronizing two spatially separated clocks,
which sets a universal limit in a measurement of remote time. If the
clock at $y$ is considered standard (zero-uncertainty), then the
remote clock at $\left|x-y\right|$ is inevitably seen uncertain.
In a homogeneous, isotropic, flat and empty space, considering a standard
clock with reading $X_{1}(y)$ is transported from place $y$ to $x$,
then the wavefunction that one finds the clock at the remote place
$x$ with reading $X_{1}(x)$ is given by \begin{equation}
\int_{X_{1}(y)}^{X_{1}(x)}\mathcal{D}X_{1}e^{-S_{1}[X_{1}]}=\frac{V^{2}}{4\pi^{2}\left|x-y\right|^{2}}e^{-2V\frac{\left[X_{1}(x)-X_{1}(y)\right]^{2}}{\left|x-y\right|}}=\frac{1}{\sigma^{4}(2\pi)^{2}}e^{-\frac{4\left[X_{1}(x)-X_{1}(y)\right]^{2}}{2\sigma^{2}}},\end{equation}
where $\int\mathcal{D}X_{1}$ is Feynman's path integral of the physical
clock. The spatial evolution of the clock broadens the wavefunction
from a standard clock (delta distribution) to a wavefunction with
finite width. The width $\sigma^{2}$ describes the uncertainty of
the reading $X_{1}(x)$ of the remote clock at $x$ with respect to
the standard clock at $y$, which is given by\begin{equation}
\sigma^{2}=\left\langle \delta X_{1}^{2}(x-y)\right\rangle =\frac{1}{V}\left|x-y\right|,\label{eq:time-fluctuation}\end{equation}
where $V$ is a 3-volume IR cut-off. Therefore, a remote simultaneity
defined by the physical clock $\left\langle X_{1}\right\rangle =\mathrm{const}$
has an intrinsic uncertainty proportional to the distance between
the remote clock and the observer. The distance dependence of the
clock uncertainty is important when the 3-volume is not infinity.
Because the IR cut-off 3-volume $V$ is as large as the cosmic scale,
the uncertainty of simultaneity can be ignored in our ordinary observation,
while it is significant when the spatial interval $\left|x-y\right|$
is also at cosmic scale. By dimensional consideration, the remote
time/simultaneity uncertainty can be written as\begin{equation}
\left\langle \delta t^{2}\right\rangle \sim L_{H}^{-3}L_{P}^{4}\left|x-y\right|,\label{eq:time-uncertain}\end{equation}
where $L_{H}\sim V^{1/3}$ and $L_{P}$ are the IR and UV cut-offs
chosen as the Hubble and Planck scale. In general, if we consider
the time is measured by a quantum physical clock, but a global parameter,
an intrinsic quantum uncertainty of remote simultaneity is inevitable.
There are two important points to emphasize: (1) the effect is different
from the time dilation, it does not change the central value $\left\langle t\right\rangle $
of the remote time, it only makes the time fuzzy with a non-vanishing
$\left\langle \delta t^{2}\right\rangle $. (2) Different from those
time effects predicted from relativity, in which time are different
in different reference frames or in a curved space, here, the effect
even happens in one reference frame and/or in a flat space. This quantum
effect that a remote clock must be uncertain Eq.(\ref{eq:time-uncertain})
provides a new explanation to the observed dark energy, the density
of which can be roughly estimated: of order $L_{H}^{-3}\sqrt{\left\langle \delta t^{2}\right\rangle ^{-1}}\sim\mathcal{O}(L_{P}^{-2}L_{H}^{-2})$,
if one considers $\left|x-y\right|\sim\mathcal{O}(L_{H})$.

Clear, the farther the distance, the weaker the clocks' correlation,
the more uncertain the time or simultaneity, so the larger the energy
fluctuations seen by the observer distance separated. It looks like
there is an apparent standard deviation of energies emerging out of
the void. The deviation or uncertainty the observer feels from the
remote clock introduces a remote energy uncertainty, according to
the uncertainty principle.

To describe this phenomenon quantitatively, we can find from Eq.(\ref{eq:action-clock}),
the observer feels energy fluctuations in a 4-volume element,\begin{align}
\left\langle \delta E_{2}(x)\delta E_{2}(0)\right\rangle d^{4}x & =\frac{\delta^{2}S_{eff}}{\delta X_{1}(x)\delta X_{1}(0)}d^{4}x\nonumber \\
 & \approx\frac{\delta^{2}S}{\delta X_{1}(x)\delta X_{1}(0)}d^{4}x=\partial_{x}^{2}\delta^{4}(x)d^{4}x.\end{align}
We have written the $S_{eff}$ by using the classical action $S$
at tree level approximation, so the leading contribution to the energy
fluctuations is expressed in terms of a widthless Dirac delta function,
while it actually has a non-vanishing width. The calculation can be
regulated when we first rewrite the Dirac delta function as a limit
of a Gaussian distribution, performing the derivatives and finally
taking the zero width limit of the Gaussian distribution back to the
delta distribution,\begin{equation}
\left\langle \delta E_{2}(x)\delta E_{2}(0)\right\rangle d^{4}x\approx\lim_{a\rightarrow0}\partial_{x}^{2}\left(\frac{1}{a^{4}\pi^{2}}e^{-\frac{4x^{2}}{a^{2}}}\right)d^{4}x=64a^{-4}\left|x-0\right|^{2}\delta^{4}(x)d^{4}x,\end{equation}
where $a$ is the UV cut-off. To regulate the result, both UV and
IR cut-offs are needed, a natural UV cut-off is the Planck length
$a=L_{P}$. Since the formula is proportional to $\left|x-0\right|^{2}$,
the fluctuations become important when the IR cut-off is at cosmic
scale, a natural choice is the Hubble scale $\left|x-0\right|=L_{H}$
as the cosmic horizon. Let us keep the squared norm $\left|x-0\right|^{2}=L_{H}^{2}$
fix and integrate over $x$, then the fluctuation of the total energy
of a Hubble scale volume is given by\begin{equation}
\left\langle \delta E_{2}^{2}\right\rangle \approx64\int d^{4}xL_{P}^{-4}L_{H}^{2}\delta^{4}(x)=64L_{P}^{-4}L_{H}^{2}.\label{eq:total-energy}\end{equation}
The proportional to the horizon area $L_{H}^{2}$ of the result known
as the area scaling is a generic feature argued by many literature
\cite{PhysRevD.34.373,PhysRevLett.71.666,PhysRevD.65.105013}. The
physical reason is transparent, since the unobservability of the total
energy of the systems inside and outside the Hubble volume, the Wheeler-DeWitt
equation provides that it is zero, then the fluctuations of their
total energies have the relation \begin{equation}
\left\langle \delta E_{inside}^{2}\right\rangle =\left\langle \delta E_{outside}^{2}\right\rangle .\end{equation}
This relation can also be proved mathematically. On the other hand,
because the inside and outside systems only share an identical bounding
surface, the relation suggests that either dispersion are proportional
to the area of the surface which scales as $L_{H}^{2}$.

We thus find a non-zero total vacuum energy fluctuation in a Hubble
scale volume (the 3-ball with fixed radius $\left|x-0\right|=L_{H}$)
Eq.(\ref{eq:total-energy}), so the averaged vacuum energy density
is\begin{equation}
\rho_{\Lambda}=\frac{\sqrt{\left\langle \delta E_{2}^{2}\right\rangle }}{\frac{4\pi}{3}L_{H}^{3}}\approx\frac{6}{\pi}L_{P}^{-2}L_{H}^{-2}\propto\frac{H^{2}}{G},\end{equation}
where $H$ is the Hubble constant, and $G$ is the Newton gravitational
constant. This result gives a correct order to the observed energy
density driving the accelerating expansion of the universe. The numerical
proportional coefficient depends on the precise nature of the cut-off,
so we would better not predict the precise value of $\rho_{\Lambda}$
unless the factors are cooked up. If I must do so, I would rather
guess $L_{P}^{2}=8\pi G$, $L_{H}^{-1}=H$, then it predicts the fraction
within the critical energy density $\rho_{c}=\frac{3H^{2}}{8\pi G}$
is $\Omega_{\Lambda}=\frac{\rho_{\Lambda}}{\rho_{c}}=\frac{2}{\pi}\approx0.637$,
which is consistent with current cosmic observation.

The key new features arising from the re-interpretation of time shown
in this section are as follows. Now the vacuum expectation value of
the energy density related to one state is vanished, however, the
energy fluctuation related to two states in the vacuum is the leading
order contribution to the gravitational effect \cite{padmanabhan1987response,2005CQGra..22L.107P},
which is a pure quantum effect originated from the intrinsic uncertainty
of the physical clock simultaneity Eq.(\ref{eq:time-fluctuation}).
These consequences provide a solution to the cosmological constant
problem. First, it explains why the conventionally predicted zero-point
energy has no gravitational effect, since the parameter time is unobservable
in the universe, so is the zero-point energy corresponding to it.
Second, it gives a prediction with correct order. 

Further more, it leads to a third consequence, it answers the {}``cosmic
coincidence'' problem, because the effective vacuum energy density
is an apparent effect due to the intrinsic fluctuations of remote
clocks but the real vacuum energy of the region-2 (like the Lorentz
contraction, which is only a visual effect but a real contraction
by a force). Note that the time $X_{1}$ here is a local internal
clock time, the functional derivative of $\rho_{\Lambda}$, $\Omega_{\Lambda}$
and the Hubble constant $H=L_{P}^{-1}$ w.r.t. the clock time $X_{1}$
vanishes, so in this sense, they are really constants and do not vary
with time. The $\rho_{\Lambda}$ is always comparable with the critical
density $\rho_{c}$ and matter density $\rho_{M}$. In a flat universe
$\Omega_{K}=0$, the fraction of the matter density is then always
seen $\Omega_{M}\approx1-\Omega_{\Lambda}$ by an internal observer
at any epoch.

At first glance, it seems that the statement {}``always comparable''
contradicts the standard picture in which the {}``dark energy''
is constant while matter are gradually diluting. How could these two
statements are both true, please do not immediately make an arbitrary
judgment that it must be wrong. In fact, the expansion of the universe
is a relative concept but absolute. The key is again that we are using
a {}``local internal clock time'' in the framework, but an {}``absolute
external time''. These two kinds of times predict different situations.
We consider the universe is divided into two parts, one is a finite
regime A in which an observer lives, and the regime B is the rest
of the universe. The notion {}``now'' in principle is a limit of
regime A shrinking to infinitely small, but in practice the regime
can be considered finite, it is the notion {}``near now'' or {}``a
near epoch''. The change in the regime B is defined relative to the
clock in regime A who is an external observer (w.r.t. regime B). While
the change in the regime A is relative to the clock also in regime
A who is an internal observer (w.r.t. regime A). As a result, the
internal observers do not see any density change of regime A with
respect to their internal clocks, although it is expanding seen by
an external observer. Because we as observers always live in the regime
A (the {}``near now'' regime), although we as external observers
can see changes in regime B, we as internal observers cannot see any
matter diluting in the regime A, since our rulers and clocks expand
accordingly, in this sense, the regime A seems like an expansion {}``static''
regime. That is the reason we always see that the matter density does
not vary with internal time and is always comparable with the apparent
{}``dark energy''.

It is worth emphasizing that {}``always comparable'' does not mean
these two as real components of the universe would be scaled in the
same way under expansion, since it is impossible to be consistent
with many observations such as the growth of large scale structure.
However, the cosmic acceleration in fact is an apparent (quantum)
cosmic variance, no matter in any epoch an (internal) observation
is performed, the mirage {}``dark energy'' is always seen being
of the order of the matter density. The evolution of the observable
universe gives place to the evolution with redshift. If one considers
the fraction of matter density evolves as $\Omega_{M}(1+z)^{3}$ from
now (could be any epoch) to a relative redshift $z$, then an internal
observer at any epoch, always {}``sees'' the vacuum energy density
become comparable with the matter density at a relatively small redshift
$z_{c}\approx\left(\frac{\Omega_{\Lambda}}{1-\Omega_{\Lambda}}\right)^{1/3}-1\approx0.3$.
{}``Why now'' seems to be a problem because of the breaking of external
time translation invariance of the scale factor, however, here the
scale factor $a\sim\partial X$ (see next section) is invariant under
internal time $X$ translation, the standard meaning that the scale
factor evolves with time is lost. In the standard external observer's
interpretation, it is a problem {}``why now'', but in a local internal
observer's view, the densities do not vary with their clocks, and
the coincident redshift $z_{c}$ is always relatively small. All in
all, the resolution of the coincidence problem is not dynamical, it
is again a consequence of using the local internal clock time.

\section{Generalization-Quantum Reference Frame}

To our knowledge, although the quantum field theories on a flat and/or
a curved spacetime background are achieved, it is difficult to treat
a quantum field system on a dynamical fluctuated spacetime background.
This is in analog with that the hydrogen atom had been explained by
the quantum mechanics, but people had no idea to the Lamb shift and
other effects due to the intrinsic fluctuations of the quantum electrodynamics
background. The status is deeply rooted in the fact that we do not
have a consistent quantum theory of spacetime. The operational re-interpretation
of time can be generalized to an operational definition of a quantum
reference frame, in which the notion of time and space coordinates
could be put on an equal footing. Since in the spirit of relativity,
the spacetime itself is nothing but the property of metric operationally
measured by physical instruments (described by quantum mechanics).
The idea provides a promising approach to treat a system on a quantum
fluctuated spacetime background. Considering reference frame (scalar)
coordinate fields $X_{\mu}(x)$ ($\mu=0,1,2,3$) defined on a flat
fixed parameter background $x_{i}$ with metric $\eta_{ij}$ ($i,j=0,...,d-1$),
and a quantum field shared the parameter background is written as
$\varphi(x)$. They are considered independent and the system is separable.
These two fields $\varphi$ and $X_{\mu}$ are defined on the parameter
background, but the parameter background does not necessarily has
any physical meaning, the important thing is the relation between
$\varphi$ and the quantum reference frame system $X_{\mu}$. At classical
level, it means the action is a functional of $\varphi$ and $\frac{\delta\varphi}{\delta X_{\mu}}$,
but at quantum level, it means that they can be written as a separable
system,\begin{equation}
S[\varphi,X_{\mu}]=\int d^{d}x\left[\frac{1}{2}\eta^{ij}\partial_{i}\varphi\partial_{j}\varphi-V[\varphi]+\frac{\lambda}{2}g_{\mu\nu}\eta^{ij}\partial_{i}X_{\mu}\partial_{j}X_{\nu}\right],\label{eq:quantum-reference-frame}\end{equation}
in which the first two terms are the actions of the field $\varphi$,
the third term describes the reference frame fields, the $g_{\mu\nu}$
is the metric of the frame manifold, i.e. $g^{\mu\nu}=\left\langle \eta^{ij}\partial_{i}X_{\mu}\partial_{j}X_{\nu}\right\rangle $,
and we have already effectively written a cosmological constant $\lambda$
in front of the reference frame term (can be viewed as a ($d-1$)-volume
averaged renormalized mass of the reference frame fields, i.e. $\lambda=m/V_{d-1}$).
Obviously, the action is formally invariant under the frame coordinates
transformation $X_{\mu}^{\prime}(x)=\omega_{\mu}^{\nu}X_{\nu}(x)+b_{\mu}$.
Note that if we do not presuppose the mean field metric $g^{\mu\nu}=\left\langle \eta^{ij}\partial_{i}X_{\mu}\partial_{j}X_{\nu}\right\rangle $,
but write it explicitly in terms of vierbein $e_{i}^{\mu}=\partial_{i}X_{\mu}$,
the precise full action is highly nonlinear.

By using the mean field approximation, it is easy to verify that this
action Eq.(\ref{eq:quantum-reference-frame}) can be reduced to our
familiar form that a quantum field lives on a curved background, \begin{align}
S_{eff} & =\int d^{4}X\left\Vert \frac{\partial x}{\partial X}\right\Vert \left[\frac{1}{4}\left(g_{\mu\nu}\eta^{ij}\partial_{i}X_{\mu}\partial_{j}X_{\nu}\right)\left(\frac{1}{2}g^{\mu\nu}\frac{\delta\varphi}{\delta X_{\mu}}\frac{\delta\varphi}{\delta X_{\nu}}+\lambda\right)-V[\varphi]\right]\\
 & =\int d^{4}X\sqrt{\det g}\left[\frac{1}{4}\mathcal{N}\left(\frac{1}{2}g^{\mu\nu}\frac{\delta\varphi}{\delta X_{\mu}}\frac{\delta\varphi}{\delta X_{\nu}}+\lambda\right)-V[\varphi]\right],\label{eq:quantum-reference-frame-variation}\end{align}
where $\mathcal{N}=\left\langle g_{\mu\nu}\eta^{ij}\partial_{i}X_{\mu}\partial_{j}X_{\nu}\right\rangle _{MF}$
is a constant calculated from the mean field value of $X_{\mu}(x)$.
If you note $\mathcal{N}=\left\langle g_{\mu\nu}g^{\mu\nu}\right\rangle _{MF}$,
it is in fact a topological invariant related to the dimension of
the reference frame. The formula Eq.(\ref{eq:quantum-reference-frame-variation})
is a generalization of Eq.(\ref{eq:effective-action-matter+clock}).
The Jacobian determinant $\left\Vert \frac{\partial x}{\partial X}\right\Vert $
requires the metric being a square matrix, thus leading to the dimensions
of the parameter space is equal to that of the reference frame fields,
i.e. $d=4$. If we do not demand such semi-classical limit, $d$ will
not necessarily be 4. A possible implication of this fact is that
a topological non-classifiable manifold at certain dimensions may
be able to reformulate as a topological classifiable manifold in other
dimensions at quantum level. It provides an alternative route to the
non-classifiable problem \cite{Hartle:1986uk,ambjorn1997quantum,kauffman1997possible}
of quantum gravity in 4-dimensions, in which the ergodicity does not
require to be abandoned.

Therefore, at classical or action level, it demonstrates an equivalence
(up to a cosmological constant $\lambda$) between the quantum reference
frame theory Eq.(\ref{eq:quantum-reference-frame}) and a quantum
field theory on a generic spacetime background. It is interesting
to note that renormalization of $\lambda$ also gives rise to a Ricci
curvature $R(g)$ of the frame manifold, $\frac{d}{d\ln k}\lambda=\frac{1}{2}Rk^{2}$.
The emerged Ricci curvature term describes the low energy dynamics
of the quantum reference frame,\begin{equation}
S_{eff}=\int d^{4}X\sqrt{\det g}\left[\frac{1}{4}\mathcal{N}\left(\frac{1}{2}g^{\mu\nu}\frac{\delta\varphi}{\delta X_{\mu}}\frac{\delta\varphi}{\delta X_{\nu}}+\lambda+\frac{1}{2}R(g)L_{P}^{-2}\right)-V[\varphi]\right].\label{eq:Ricci}\end{equation}

It is striking that Einstein's theory of gravity emerges as a low
energy effective quantum dynamics of the reference frame, the (relational)
quantum reference frame theory itself automatically contains a theory
of gravity. Although we have shown the classical equivalence between
the theory Eq.(\ref{eq:quantum-reference-frame}) and conventional
dynamical spacetime theory, at quantum level these two theories are
very different. First, the theory Eq.(\ref{eq:quantum-reference-frame})
relates to a Wheeler-DeWitt equation, and the states defined on the
hypersurface of parameter background $x$ of the equation are entangled
states, entangling the state of quantum field $\varphi$ with the
state of the reference frame $X_{\mu}$, only the relational interpretation
of these states is reasonable, the entangled state suggests that the
theory is a parameter background independent theory. In contrast,
the theory Eq.(\ref{eq:quantum-reference-frame-variation}) relates
to an approximate Schrodinger equation, the state of it is thought
defined on the hypersurface of $X_{\mu}$, which can only be realized
when the field $X_{\mu}$ are treated semi-classically, only in this
case, the theory has a standard absolute probability interpretation.
Second, there is no zero-point energy if you stand on the quantum
reference frame, since the reference frame is also fluctuating at
quantum level. Third, the most important feature is that, Eq.(\ref{eq:quantum-reference-frame})
has a well-defined quantum theory defined on the flat parameter background
$x$, while it is difficult to treat Eq.(\ref{eq:quantum-reference-frame-variation})
and/or Eq.(\ref{eq:Ricci}) quantum mechanically when $X_{\mu}$ is
a dynamical background spacetime. In this sense, Eq.(\ref{eq:quantum-reference-frame})
may be a good starting point to study a quantum theory of gravity.

\section{Conclusions}

In this paper, we abandon the interpretation that time is a global
parameter in quantum mechanics, replace it by a quantum dynamical
variable playing the role of time. The operational re-interpretation
of time causes a new notion of energy and important consequences.
We find (1) the expectation value of the zero-point energy under the
new time variable vanishes; (2) the leading contribution to the gravitational
effect is the energy fluctuation, the vacuum energy fluctuation effectively
gives a correct order to the observed {}``dark energy'', $\rho_{\Lambda}=\frac{6}{\pi}L_{P}^{-2}L_{H}^{-2}$;
(3) The vacuum energy density is always comparable with the matter
energy density seen by an observer using the local internal clock
time. The three of the consequences from the time re-interpretation
provides a solution to the cosmological constant problem.

The re-interpretation of time also leads to several conceptual consequences.
(1) The new quantum time variable is able to reduce to conventional
parameter time as a limit of semi-classical approximation. (2) The
Wheeler-DeWitt equation plays a more fundamental role than the textbook
Schrodinger equation. The Schrodinger equation is a derivation under
the semi-classical approximation of the Wheeler-DeWitt equation. (3)
The solution of the Wheeler-DeWitt equation is in general an entangled
state, which leads to the consequence that the absolute probability
interpretation of the textbook quantum mechanics is required to be
replaced by a relational interpretation with the help of the joint
probability. (4) The entangled state solution of the Wheeler-DeWitt
equation implies that not only the to-be-measured system but also
the measuring instruments (such as the clock) both are required to
be described by the quantum mechanics.

The idea of re-interpretation of time can be generalized to a more
general version of a quantum reference frame, in which we could put
the space and time on an equal footing. This framework provides us
a new approach to treat the spacetime quantum dynamically, and leads
to a possible route to the non-classifiable problem of quantum gravity
in 4-dimensions.
\begin{acknowledgments}
This work was supported in part by the National Science Foundation
of China (NSFC) under Grant No.11205149.
\end{acknowledgments}
\bibliographystyle{apsrev}

\end{document}